\documentclass[letterpaper, 10 pt, conference]{ieeeconf}  
\usepackage{amsmath,amsfonts}
\usepackage{algorithmic}
\usepackage{algorithm}
\usepackage{array}
\usepackage{textcomp}
\usepackage{stfloats}
\usepackage{url}
\usepackage{verbatim}
\usepackage{graphicx}
\usepackage{graphbox}
\usepackage{cite}
\usepackage{subcaption}
\usepackage{xcolor}
\usepackage{hyperref}

\newcommand{\edits}[1]{\textcolor{black}{#1}}

\DeclareMathOperator*{\argmin}{argmin}

\IEEEoverridecommandlockouts                              

\title{\LARGE \bf
Control-Coherent Koopman Modeling: A
Physical Modeling Approach
}

\author{H. Harry Asada$^{1}$, \textit{Life Fellow} and Jose A. Solano-Castellanos$^{1}$
\thanks{$^{1}$Authors are members of the d'Arbeloff Laboratory in the Department of Mechanical Engineering at the Massachusetts Institute of Technology, Cambridge, MA, 02139, USA. {\tt\small \{asada, jsolanoc\}@mit.edu}. This material is based upon work supported by the National Science Foundation under Grant No. NSF-CMMI 2021625.}
}

\begin{document}

\maketitle

\begin{abstract}
The modeling of nonlinear dynamics based on Koopman operator theory, originally applicable only to autonomous systems with no control, is extended to non-autonomous control system without approximation \edits{of} the input matrix. Prevailing methods using a least square estimate of the input matrix may result in an erroneous input matrix, misinforming the controller. Here, a new method for constructing a Koopman model that \edits{yields} the exact input matrix is presented. A set of state variables are introduced so that the control inputs are linearly involved in the dynamics of actuators. With these variables, a lifted linear model with the exact input matrix, called a Control-Coherent Koopman Model, is constructed by superposing control input terms, which are linear in local actuator dynamics, to the Koopman operator of the associated autonomous nonlinear system.  As an example, the proposed method is applied to multi degree-of-freedom robotic arms, which are controlled with Model Predictive Control \edits{(MPC)}. It is demonstrated that the prevailing Dynamic Mode Decomposition with Control (DMDc) using an approximate input matrix does not provide a satisfactory result, while the Control-Coherent Koopman Model performs well with the correct input matrix, even performing better than the bilinear formulation of the Koopman operator.
\end{abstract}



\section{INTRODUCTION}

Koopman Operator theory has the potential to be a breakthrough in representation of complex nonlinear dynamics. A globally linear, unified representation facilitates control synthesis and analysis. Powerful linear systems theory and techniques can be applied to complex nonlinear systems. It has already had significant impacts upon various branches of control theory and applications, ranging from system identification \cite{Mauroy2020Koopman-BasedIdentification}, Model Predictive Control \cite{Korda2018LinearControl}, and robust control \cite{Han2022DeSKO:Operator} to soft robot modeling and control \cite{Bruder2021Data-DrivenTheory}, vehicle control \cite{Cibulka2020ModelOperator}, and active robot learning \cite{Abraham2019ActiveOperators}. 

Despite promising reports, a fundamental problem has not yet been solved. Is Koopman operator theory applicable to non-autonomous systems with control? The Koopman Operator theory is originally applicable only to autonomous systems having no exogenous input \cite{Koopman1931HamiltonianSpace, ROWLEY2009SpectralFlows}. Almost all control systems are non-autonomous with control, to which the Koopman theory does not apply in the strict sense. Criticisms of the Koopman Operator approach often pertain to this limitation.

The Koopman research community has been attempting to remove this limitation. An ad hoc method is to simply approximate the control input term to a linear term with constant coefficients, $\dot{z} = Az + Bu$, where $z$ is the lifted state and $u$ is the input. Matrices $A$ and $B$ are obtained from least square estimation. The \edits{commonly used} method referred to as Dynamic Mode Decomposition with Control (DMDc) is in this category \cite{Proctor2016DynamicControl}. Assuming a constant input matrix $B$ is difficult to justify because the coefficients are state dependent in many nonlinear systems. Although state variables can be lifted for global linearization, control variables cannot, because the number of independent input variables is physically determined. 

The second method is to treat control inputs as part of the independent state variables and apply the standard Koopman Operator to the augmented state variables, $[x^T, \ u^T]^T$\cite{Proctor2018GeneralizingControl}. This entails a prescribed differential equation governing the time evolution of control $u$. This method cannot be used for designing a controller from a model because the control is determined before the Koopman model is obtained. This formulation results in a causality violation in most control design settings. 

A more rigorous and more accurate method is to use a bilinear formulation. Due to the state-dependent nature of control input terms, it is difficult to approximate it to linear terms. Instead, these can be more accurately approximated to bilinear terms, where the control input terms are modeled as products of state variables and control variables \cite{Bruder2021AdvantagesDynamicsb}. This bilinear approximation provides \edits{a} more accurate approximation, \edits{but the resultant bilinear Koopman model is more complex compared to the completely linear Koopman model}. More recently, another approximation method and its error bound have been presented, where the \edits{input} matrix is still not globally constant \cite{Iacob2024KoopmanInputs}. 

Here, we present a method for constructing a Koopman operator for a class of control systems without approximation of the input matrix. Integrity and coherency of the input matrix are crucial for proper control design. An input matrix that is determined merely by curve fitting to data may not have the right structure, which may misinform control design. The proposed method guarantees the coherent, correct structure by construction. No curve fitting to a linear or bilinear parametric model is used. The method is based on causality of physical system modeling applied to actuator power-train dynamics. The new method will fill the theoretical and technical gap between the Koopman operator theory and what is needed in control engineering. The method is applicable to a vast number of control systems. 

The outline of the manuscript is the following: Section \ref{sec:Background} presents the preliminaries and problem formulation; Section \ref{sec:Control-Coherent} delves into the developement of the Control-Coherent Koopman (CCK) approach, the main contribution of this paper. Section \ref{sec:RobotDynamics} illustrates how the CCK approach can be implemented to the dynamics of a robotic manipulator. Finally, Section \ref{sec:Simulation} presents numerical results. In particular, we will show the effectiveness of our proposed method in controlling a two-degree-of-freedom robotic manipulator. We compare the tracking performance of the CCK model against the standard DMDc and a more rigorous bilinear model as described in \cite{Bruder2021AdvantagesDynamicsb}, where our approach is able to outperform both DMDc and the bilinear model in all the tracked trajectories.


\section{BACKGROUND AND PROBLEM FORMULATION}\label{sec:Background}

Consider a discrete-time, nonlinear dynamical system given by
\begin{equation}\label{eq:dynamics}
    x_{t+1} = f(x_t, u_t)
\end{equation}
where $x_t \in \mathcal{X} \subset \mathbb{R}^n$ is the independent state, $u_t \in \mathcal{U} \subset \mathbb{R}^r$ is the input, and $f$ is a continuously differentiable function, defined in compact sets $\mathcal{X}$ and $\mathcal{U}$. The autonomous system associated to (\ref{eq:dynamics}) is
\begin{equation}
    x_{t+1} = F(x_t)
\end{equation}
where $u_t \equiv 0$ and the function $F$ is a self-map, $F: \mathcal{X}\rightarrow\mathcal{X}$.

Let $\{g_i(x_t)\}^\infty_{i=1}$ be observables that span a Hilbert space $\mathcal{H}$. Assume that the observables compositional with the self-map $F$
are involved in the Hilbert space,
\begin{equation}\label{eq:observables}
    g_i \circ F \in \mathcal{H}, \quad i = 1, 2, \cdots.
\end{equation}
Then, the following Koopman operator $A$ exists:
\begin{equation}\label{eq:KoopmanAuto}
    z_{t+1} = Az_t
\end{equation}
where $z_t = [g_1(x_t), g_2(x_t), \cdots ]^T$ is the infinite-dimensional state lifted with $\{g_i\}^\infty_{i=1}$ \cite{Koopman1931HamiltonianSpace, Asada2023GlobalMethodb}.

The Koopman operator can be obtained with various methods. The most prevailing is the extended dynamic mode decomposition (EDMD), which is based on least square estimate and singular value decomposition. A more rigorous method is to obtain the Koopman operator from inner products of the observables and their composition with the self-map, state transition function $F$. Post-multiply $z_t^T$ to both sides of (\ref{eq:KoopmanAuto}) and integrate them over the dynamic range of the independent state yields
\begin{equation}\label{eq:Q}
    Q = AR
\end{equation}
where
\begin{equation}
    Q = \langle z_{t+1}, z_t \rangle = \int_{\mathcal{X}} g_i(F(x)) \cdot g_j(x) \ dx
\end{equation}
\begin{equation}
    R = \langle z_{t}, z_t \rangle = \int_{\mathcal{X}} g_i(x) \cdot g_j(x) \ dx
\end{equation}

The Koopman operator is the solution to the linear equation (\ref{eq:Q}). This method directly encodes the state transition function $F$ with a given set of observables $\{g_i\}^\infty_{i=1}$, called Koopman Direct Encoding \cite{Asada2023GlobalMethodb}.

Finding an effective set of observables is a challenge. Among others, the use of deep learning is an effective data-driven method for finding observables that can approximate the Koopman operator with a lower dimensional model \cite{Li2017ExtendedOperator, Lusch2018DeepDynamics, Yeung2019LearningSystems, Han2020DeepControl}. Those methods established in the Koopman operator theory are for autonomous systems with no control input. The objective of this paper is to develop a Koopman lifted linearization method for the non-autonomous system (\ref{eq:dynamics}) in the following form:
\begin{equation}\label{eq:Control_Coherent}
    z_{t+1} = Az_t + Bu_t
\end{equation}
which is linear in control $u_t$ with a constant input matrix.

In the DMDc method, the $A$ and $B$ matrices are given by
\begin{equation}\label{eq:DMDc}
    (\hat{A}, \hat{B}) = \argmin_{A, B} \sum_{i=1}^N \left|z_+(i) - Az_-(i) - Bu_-(i)\right|^2
\end{equation}
where the $A$ and $B$ matrices are fitted to data $\{z_+(x(i)), z_-(x(i)), u_-(i) \ | \ i = 1, \cdots, N\}$ consisting of before ($z_-,\ u_-$) and after ($z_+$) each transition. In this formulation, the approximate input matrix $\hat{B}$ is determined simply by minimizing the squared error. Note that the nonlinear dynamics with regard to the state $x_t$ can be globally linearized by lifting the state. However, it does not apply to the input $u$. This may cause an incoherent input matrix $\hat{B}$, although the curve fitting shows a good agreement. The method proposed can solve these problems.


\section{CONTROL-COHERENT KOOPMAN MODELING}\label{sec:Control-Coherent}

Control systems are activated with actuators that drive some state variables directly in response to control input. We divide the state space into the one associated to a set of state variables $p_t \in \mathcal{P} \subset \mathbb{R}^m$ that is driven directly with the input $u_t$ and the rest
of the state variables $q_t \in \edits{\mathcal{X}_q \subset} \ \mathbb{R}^{n-m}$ that are not directly driven by $u_t$ but indirectly through $p_t$.
\begin{equation}
    x_t = \left[ \begin{matrix}
        p_t \\
        q_t
    \end{matrix}\right]
\end{equation}

\textbf{Definition 1 (Actuation Subsystem):} The dynamical system (\ref{eq:dynamics}) is said to have an actuation subsystem if the state equation (\ref{eq:dynamics}) can be divided into the following two:
\begin{equation}\label{eq:p}
    p_{t+1} = f_p(x_t,u_t)
\end{equation}
\begin{equation}\label{eq:q}
    q_{t+1} = f_q(x_t)
\end{equation}
where $f_q: \mathcal{X} \rightarrow \mathcal{X}_q$ and $f_p: \mathcal{X} \times \mathcal{U} \rightarrow \mathcal{P}$ are continuously differentiable, and $u_t$ is involved in each component $f_{p,i}$,
\begin{equation}
    \frac{\partial f_{p,i}}{\partial u_t}  \neq 0, \ i = 1, \cdots, m
\end{equation}

In the following derivation, we are interested in an actuator subsystem that is nonlinear in state, $x_t$, but is linear in input, $u_t$.

\textbf{Definition 2 (Linear Actuation):} If the actuation subsystem involved in the dynamical system (\ref{eq:dynamics}) is in the following form
\begin{equation}\label{eq:linear_actuation}
    p_{t+1} = h(x_t) + B_p u_t
\end{equation}
the system is said to be linear in actuation.

\textbf{Remark 1}: As will be discussed further in the following examples, actuation subsystems are linear in control in most electro-mechanical systems. A DC motor and a brushless DC motor, for example, have equations of motion given by:
\begin{equation}
    I \Ddot{\phi} = \tau_m - \tau_{load}
\end{equation}
where $\phi$ is rotor displacement, $I$ is the rotor inertia, $\tau_m$ is the torque generated by the actuator, and $\tau_{load}$ is the load torque as the actuator is engaged with a load. Treating $\tau_m$ as input yields an actuator subsystem that is linear in actuation.

\textbf{Remark 2}: The actuation subsystem must have independent state variables $p$. This requirement can be met in several ways. For example, if the power train connecting an actuator to its load has a compliance, the local actuator subsystem and the main system driven via the power train can possess independent state variables, $p$ and $q$. These independent state variables are dynamically coupled through, for example, the power train having a coupling impedance. 

A Control-Coherent Koopman Model (\ref{eq:Control_Coherent}) can be constructed for a dynamical system with an actuation subsystem that is linear in actuation.

\textbf{Proposition (Control-Coherent Koopman Model):} If the dynamical system (\ref{eq:dynamics}) has an actuation subsystem that is linear in actuation in the form of (\ref{eq:linear_actuation}), and if observables $\{g_i\}^\infty_{i=1}$ satisfying conditions (\ref{eq:observables}) exist and they include the state variables of the actuation subsystem, $p$, in the observables, then the Control-Coherent Koopman Model of (\ref{eq:dynamics}) is given by (\ref{eq:Control_Coherent})
where $A$ is the Koopman operator of the associated autonomous system given by (\ref{eq:q}) and (\ref{eq:p_tilde})
\begin{equation}\label{eq:p_tilde}
    \Tilde{p}_{t+1} = h(x_t)
\end{equation}
which is valid in the compact set $\mathcal{X}$, and the input matrix $B$ is given by 
\begin{equation}
    B = \left[
    \begin{matrix}
        B_p \\
        0
    \end{matrix}
    \right].
\end{equation}

\begin{proof} By construction we can show that the Control Coherent Koopman model (\ref{eq:Control_Coherent}) exists. Without loss of generality, we assume that the first $m$ observables are the state variables of the actuation subsystem.
\begin{equation}
z_t = \left[
\begin{matrix}
    p_t \\
    y_t
\end{matrix}
\right]
\end{equation}
where $p_t = \left[g_1, \ \cdots, \ g_m \right]^T$ and $y_t = \left[g_{m+1}, \ g_{m+2}, \ \cdots \right]^T$.

Then the Koopman operator for the autonomous system (\ref{eq:q}) and (\ref{eq:p_tilde}) exists and can be expressed as
\begin{equation}\label{eq:modified_koopman}
\left[
\begin{matrix}
    \tilde{p}_{t+1} \\
    y_{t+1}
\end{matrix}
\right] = 
\left[
\begin{matrix}
    A_{pp} & A_{py} \\
    A_{yp} & A_{yy}
\end{matrix}
\right] 
\left[
\begin{matrix}
    p_{t} \\
    y_{t}
\end{matrix}
\right]
\end{equation}
where the matrix and vectors are divided into blocks associated to the state of the actuator dynamics and the rest of the observables, $y$. Substituting $\tilde{p}_{t+1} = p_{t+1} - B_pu_t $ into (\ref{eq:modified_koopman}) and moving the term $B_p u_t$ to the right-hand side yields
\begin{equation}\label{eq:koopman_coherent}
    \left[
\begin{matrix}
    p _{t+1} \\
    y_{t+1}
\end{matrix}
\right] = 
\left[
\begin{matrix}
    A_{pp} & A_{py} \\
    A_{yp} & A_{yy}
\end{matrix}
\right] 
\left[
\begin{matrix}
    p_{t} \\
    y_{t}
\end{matrix}
\right] + 
\left[
\begin{matrix}
    B_p \\
    0
\end{matrix}
\right] u_t
\end{equation}
\end{proof}

This is the Control-Coherent Koopman Model of the non-autonomous, nonlinear dynamical system (\ref{eq:dynamics}).

\textbf{Remark 3:} The Koopman operator given by (\ref{eq:modified_koopman}) must be valid in the compact set $\mathcal{X}$. Especially, the dynamic range of $p$ must include all the states of $p$ that can be driven by the input $B_p u_t$.

\textbf{Remark 4:} The above state equation (\ref{eq:koopman_coherent}) manifests that the actuator input $u_t$ drives the actuator subsystem state to $p_{t+1}$ and that all others in $y$ are affected through $p_{t+1}$ at the \textit{next cycle}, $t + 2$. This causal sequence agrees with our observation and the causality analysis of physical modeling theory \cite{Karnopp2006SystemSystems}. The causality dictates that the impact of the input $u_t$ is captured and confined within the actuator dynamics in the first cycle, $t \rightarrow t +1$, before being transmitted to the dynamics of $y_t$ in the second cycle, $t + 1 \rightarrow t + 2$. If the actuator state $p_t$ is eliminated, it implies that this transmission delay is eliminated. As a result, (\ref{eq:p}) becomes algebraic and the input is directly involved in (\ref{eq:q}), which prevents the application of the Koopman operator. The actuation subsystem must possess independent state variables.

If the linearity in actuation (\ref{eq:linear_actuation}) is satisfied globally, in other words, the original dynamical system (\ref{eq:dynamics}) is linear in control, the problem is straightforward. In most nonlinear dynamical systems, however, linearity in control occurs only for actuator subsystems. Modeling the actuation subsystem with independent state variables is essential to fill the gap between the original Koopman operator theory for autonomous systems and real-world control systems.


\section{APPLICATION TO ROBOT ARM DYNAMICS}\label{sec:RobotDynamics}

The Control-Coherent Koopman Modeling method is applicable to a broad spectrum of nonlinear dynamical systems. In this section, we will demonstrate how the method is applied to practical problems.

Consider the equations of motion (EoM) of an $N$ degree-of-freedom (DoF) robot in joint coordinates $\theta \in \mathbb{R}^N$ \cite{Slotine1992RobotControl}.
\begin{equation}
    H(\theta)\Ddot{\theta} + C(\theta,\dot{\theta})\dot{\theta} + G(\theta) = \tau_j
\end{equation}
where $H(\theta)$ is an $N \times N$ inertia matrix, $C(\theta,\dot{\theta})\dot{\theta}$ is the Coriolis and centrifugal terms, $G(\theta)$ is the gravity vector, and $\tau_j \in \mathbb{R}^N$ is the joint torque vector. If we define $\theta$ and $\dot{\theta}$ to be independent state variables, the state equation can be given by
\begin{equation}\label{eq:robot_dynamics}
    \frac{d}{dt}\left[
    \begin{matrix}
        \theta \\
        \dot{\theta}
    \end{matrix}
    \right] = 
    \left[
    \begin{matrix}
        \dot{\theta} \\
        H(\theta)^{-1}\tau_j - H(\theta)^{-1}\left(C\dot{\theta} + G(\theta)\right)
    \end{matrix}
    \right]
\end{equation}

In most robotics literature, joint torques $\tau_j$ are treated as control input. In consequence, the state equation is not linear in control: the joint torques are multiplied by the inverse of the state-dependent inertia matrix, $ H(\theta)^{-1}$. The Koopman operator theory cannot be applied to this form of dynamical system.

\begin{figure}[ht]
    \centering  \includegraphics[width=0.8\columnwidth]{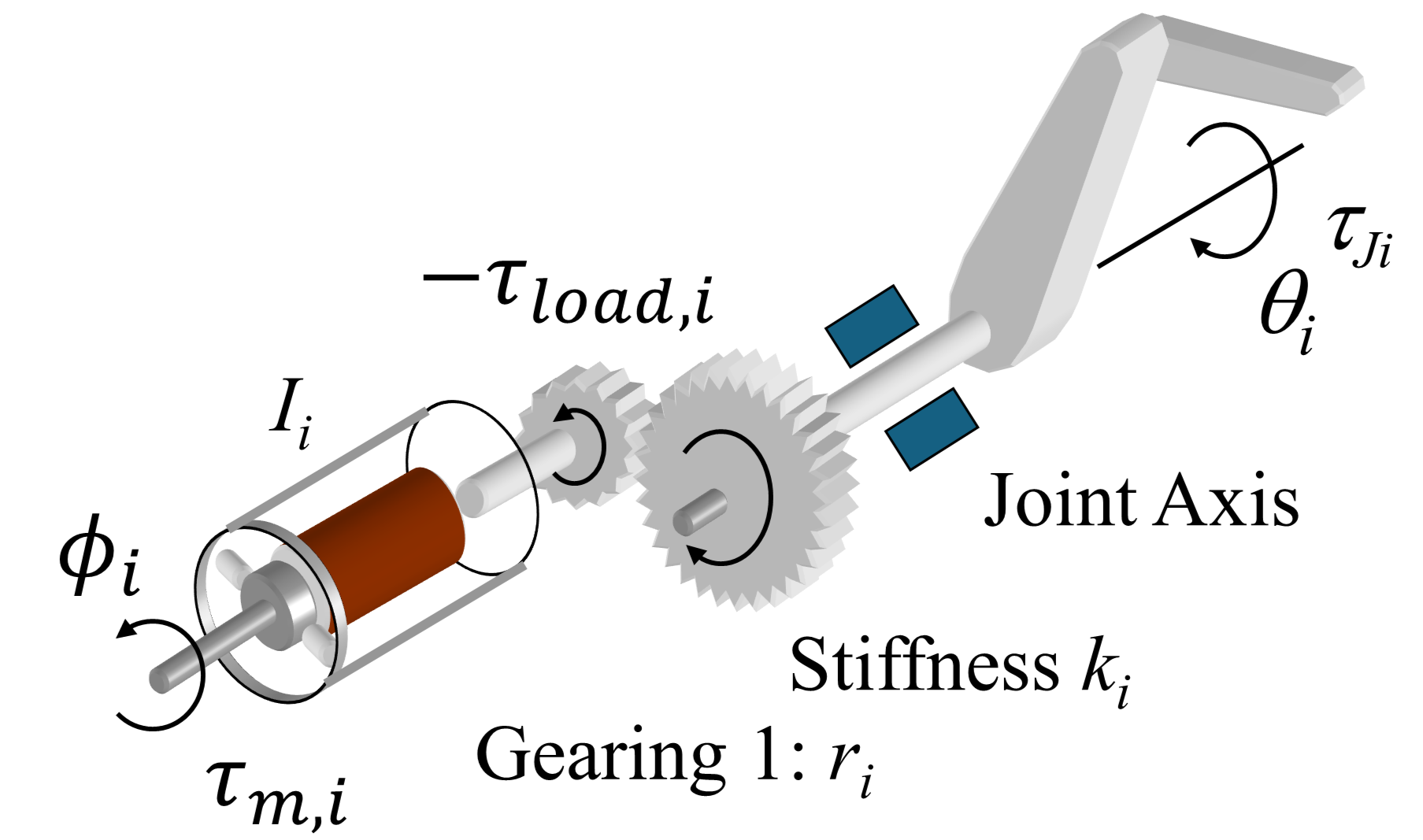}
    \caption{Dynamic modeling of the i\textsuperscript{th} joint actuator with torsional compliance at the power train.}
    \label{fig:PowerTrain}
\end{figure}

Here, we apply the Control-Coherent Koopman Modeling based on actuator dynamics. Joint torques $\tau_j$ pertain to the dynamics of actuators driving individual joints. As shown in Fig. \ref{fig:PowerTrain}, let $\phi_i$ be the rotor displacement of the actuator driving the $i^{th}$ joint, $\tau_{mi}$ the actuator torque, $b_i$ the damping constant, and $I_i$ the rotor inertia. Then, the following equation of motion is obtained,
\begin{equation}\label{eq:actuator_dynamics}
    \Ddot{\phi}_i = \frac{1}{I_i}\tau_{mi} - \frac{1}{I_i}b_i\dot{\phi}_i - \frac{1}{I_i}\tau_{load,i}, \ 1 \leq i \leq N
\end{equation}
where $\tau_{load,i}$ is the load torque from the $i^{th}$ joint. Note that the local dynamics of this actuator are linear in control. This is true for most electro-mechanical robotic systems, because the inertia of each actuator’s moving part, e.g. a motor rotor, has a constant inertia.

As shown in Fig. \ref{fig:PowerTrain}, the power train of each joint actuator, comprising a gearing and transmission mechanism, inevitably possesses some torsional compliance \cite{Spong2020RobotControl}. Suppose that a gear reducer of gear ratio $1:r_i$ connects the actuator shaft and the joint axis with a torsional stiffness $k_i$, then the joint torque $\tau_{ij}$ and the load torque $\tau_{load,i}$ are related as
\begin{equation}\label{eq:tau}
    \tau_{ji} = r_i\tau_{load,i} = r_i k_i (\phi_i - r_i \theta_i )
\end{equation}
where $\theta = \left[\theta_1, \theta_2, \cdots, \theta_N \right]^T$ and $\phi = \left[\phi_1, \phi_2, \cdots, \phi_N \right]^T$

With the compliance at the gearing, $\phi_i$ and $\theta_i$ become independent generalized coordinates. In discrete time, the independent state variables $x_t = \left[p_t^T, q_t^T \right]^T$ are those of the actuators and the arm linkage, respectively.
\begin{subequations}
\begin{equation}\label{eq:p_phi_1}
    p_t = \left[
    \begin{matrix}
        \phi_t \\
        \dot{\phi}_t
    \end{matrix}
    \right]
\end{equation}
\begin{equation}\label{eq:q_theta}
    q_t = \left[
    \begin{matrix}
        \theta_t \\
        \dot{\theta}_t
    \end{matrix}
    \right]
\end{equation}    
\end{subequations}

Using (\ref{eq:tau}) in (\ref{eq:robot_dynamics}), the state equation of the arm linkage can be approximated to the following form in discrete time.
\begin{subequations}\label{eq:discretized_theta}
    \begin{equation}
        \theta_{t+1} = \theta_t + \Delta t \dot{\theta}_t
    \end{equation}
    \begin{multline}
        \dot{\theta}_{t+1} = \dot{\theta}_t + \Delta t H(\theta_t)^{-1}\left[\mathbf{r}\mathbf{k}(\phi_t - \mathbf{r}\theta_t) - C\dot{\theta}_t - G \right]
    \end{multline}
\end{subequations}
where $\mathbf{r} = \text{diag}(r_1, \cdots, r_N)$ and $\mathbf{k} = \text{diag}(k_1, \cdots, k_N)$. Using (\ref{eq:tau}) in (\ref{eq:actuator_dynamics}) yields
\begin{subequations}\label{eq:discretized_phi}
    \begin{equation}
        \phi_{t+1} = \phi_t + \Delta t \dot{\phi}_t
    \end{equation}
    \begin{multline}
        \dot{\phi}_{t+1} = \dot{\phi}_t - \Delta t \mathbf{I}^{-1}\left[\mathbf{b}\dot{\phi}_t + \mathbf{k}(\phi_t - \mathbf{r}\theta_t) + \tau_m\right]
    \end{multline}
\end{subequations}
where $\mathbf{I} = \text{diag}(I_1, \cdots, I_N)$ and $\mathbf{b} = \text{diag}(b_1, \cdots, b_N)$. The control input term $\tau_m = \left[\tau_{m1}, \cdots, \tau_{mN} \right]^T$ are linearly involved in the actuator dynamics. Moving the linear control term to the left-hand side, the second equation becomes.
\begin{equation}\label{eq:phi_tilde}
    \dot{\tilde{\phi}}_{t+1} = \dot{\phi}_t - \Delta t \mathbf{I}^{-1}\left[\mathbf{b}\dot{\phi}_t + \mathbf{k}(\phi_t - \mathbf{r}\theta_t)\right]
\end{equation}
where $\dot{\tilde{\phi}}_{t+1} = \dot{\phi}_{t+1} - \Delta t \mathbf{I}^{-1}\tau_m$.

The Koopman operator $A$ is computed for the autonomous system, (\ref{eq:discretized_theta}) and (\ref{eq:phi_tilde}). Using the resultant $A$ matrix, the Control-Coherent Koopman Model (\ref{eq:koopman_coherent}) can be obtained by moving the control term $\Delta t \mathbf{I}^{-1} \tau_m$ to the right-hand side.

It should be noted that the actuator dynamics (\ref{eq:discretized_phi}) is linear in this model. Therefore, unlike the arm linkage dynamics (\ref{eq:discretized_theta}), which must be lifted with many observables for linearization, the actuator dynamics (\ref{eq:discretized_phi}) does not need to lift, since it is already linear. This implies that the upper block matrices in (\ref{eq:koopman_coherent}), $A_{pp}$, $A_{py}$, can be replaced by the parameters involved in (\ref{eq:discretized_phi}) and that the coefficients associated to the observables for lifting linearization are all zero.
\begin{equation}
    \left[
    \begin{matrix}
        p_{t+1} \\
        y_{t+1}
    \end{matrix}
    \right] = 
    \left[
    \begin{matrix}
        \tilde{A}_{pp} & \tilde{A}_{py} \ 0 \ \cdots \ 0 \\
        A_{yp} & A_{yy}
    \end{matrix}
    \right] 
    \left[
    \begin{matrix}
        p_{t} \\
        y_{t}
    \end{matrix}
    \right] + \left[
    \begin{matrix}
        B_p \\
        0
    \end{matrix}
    \right]u_t
    \label{eq:p_phi}
\end{equation}
\vspace{0.1mm}


\section{NUMERICAL SIMULATION: A TWO-LINK ROBOT ARM}\label{sec:Simulation}

\begin{figure}[H]
    \centering  \includegraphics[width=0.825\columnwidth]{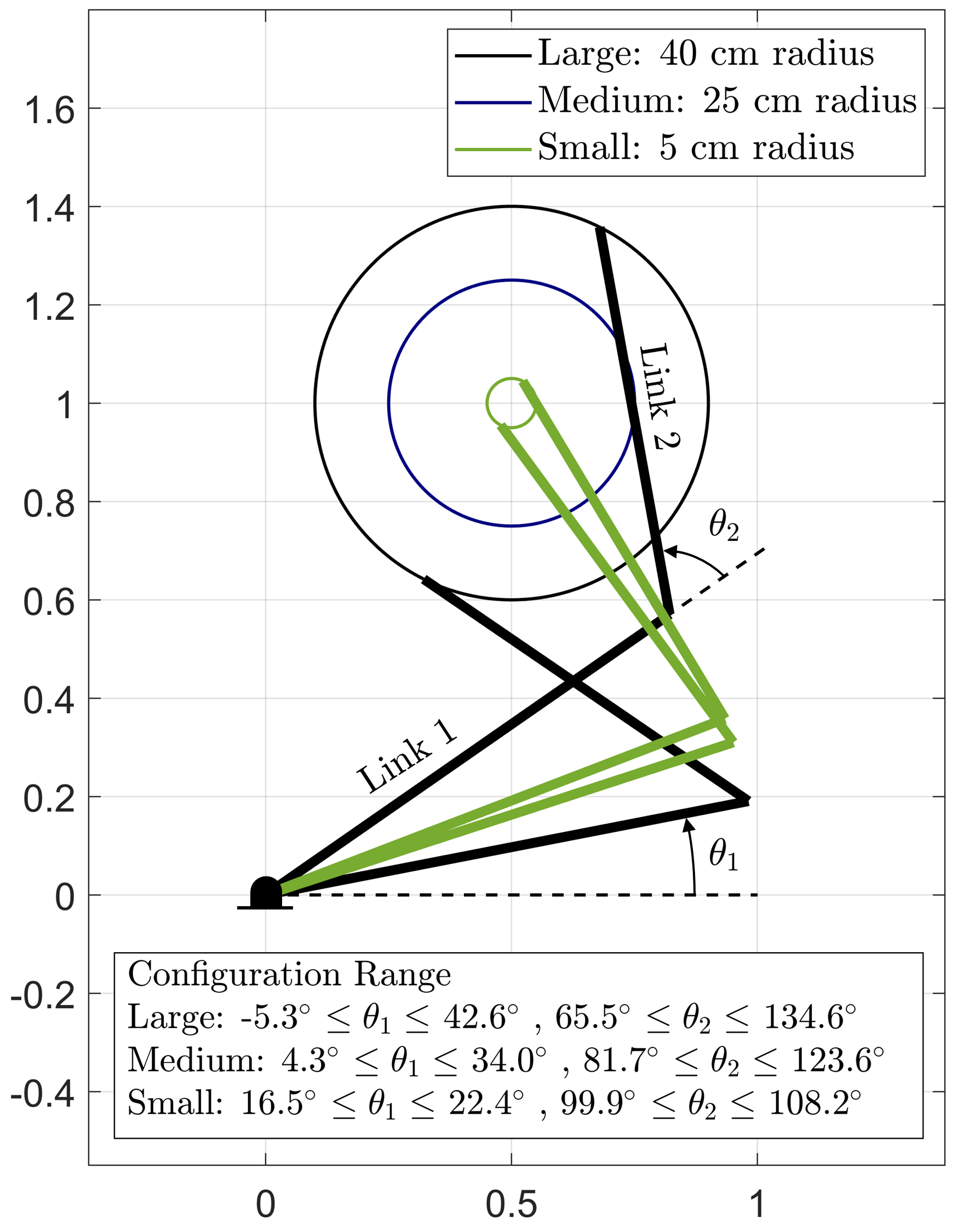}
    \caption{Configuration range for a two-degree-of-freedom robotic arm following three circular trajectories.}
    \label{fig:Configuration}
\end{figure}

The proposed Control Coherent Koopman (CCK) modeling method is implemented on a two-link robot shown in
Fig. \ref{fig:Configuration}, and Model Predictive Control (MPC) is applied to the CCK model. The first link measures $1 \ m$ and weighs $5 \ kg$, while the second link is $0.8 \ m$ and $4 \ kg$. The arm moves in a horizontal plane with no gravity. The order of the arm dynamics is four for the two links, and that of the actuator dynamics is four. Two-hundred Radial Basis Functions (RBFs) are used as observables in addition to the actuator-augmented state, where the center locations of RBFs are determined with the k-means clustering method. The $A$ matrix of the CCK model, $A_{CCK}$, is obtained with data containing both the autonomous $(u_t = 0)$ and non-autonomous $(u_t \neq 0)$ response of the system. The input matrix of the CCK model, $B_{CCK}$, is determined from the effective actuator rotor inertia as in (\ref{eq:actuator_dynamics}) and arranged as in (\ref{eq:p_phi}).

\begin{figure*}[!t]
    \centering
    \subfloat[CCK]{\includegraphics[height=5.1cm]{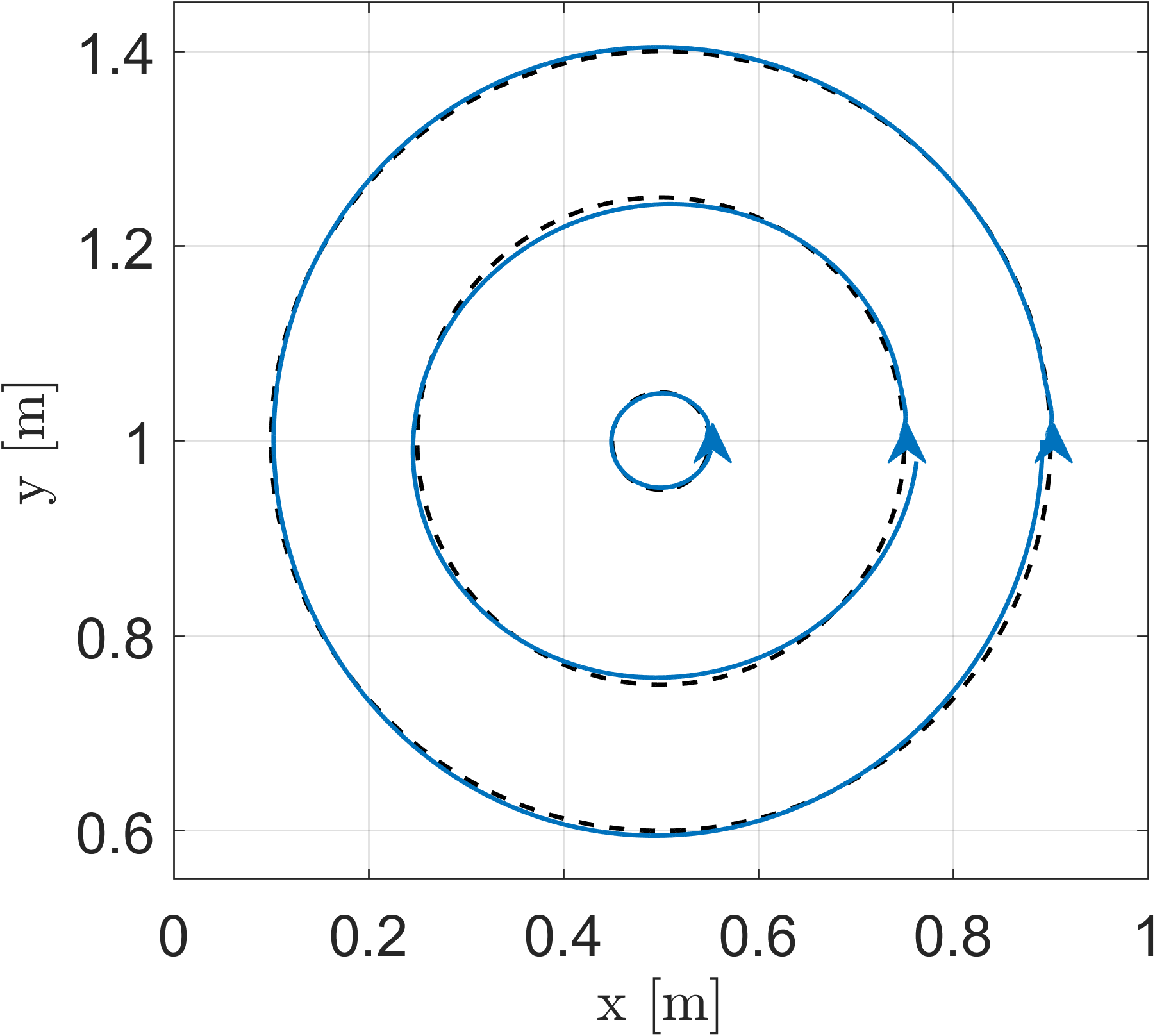}%
    \label{fig:CCKTracking}}
    \hfil
    \subfloat[DMDc]{\includegraphics[height=5.1cm]{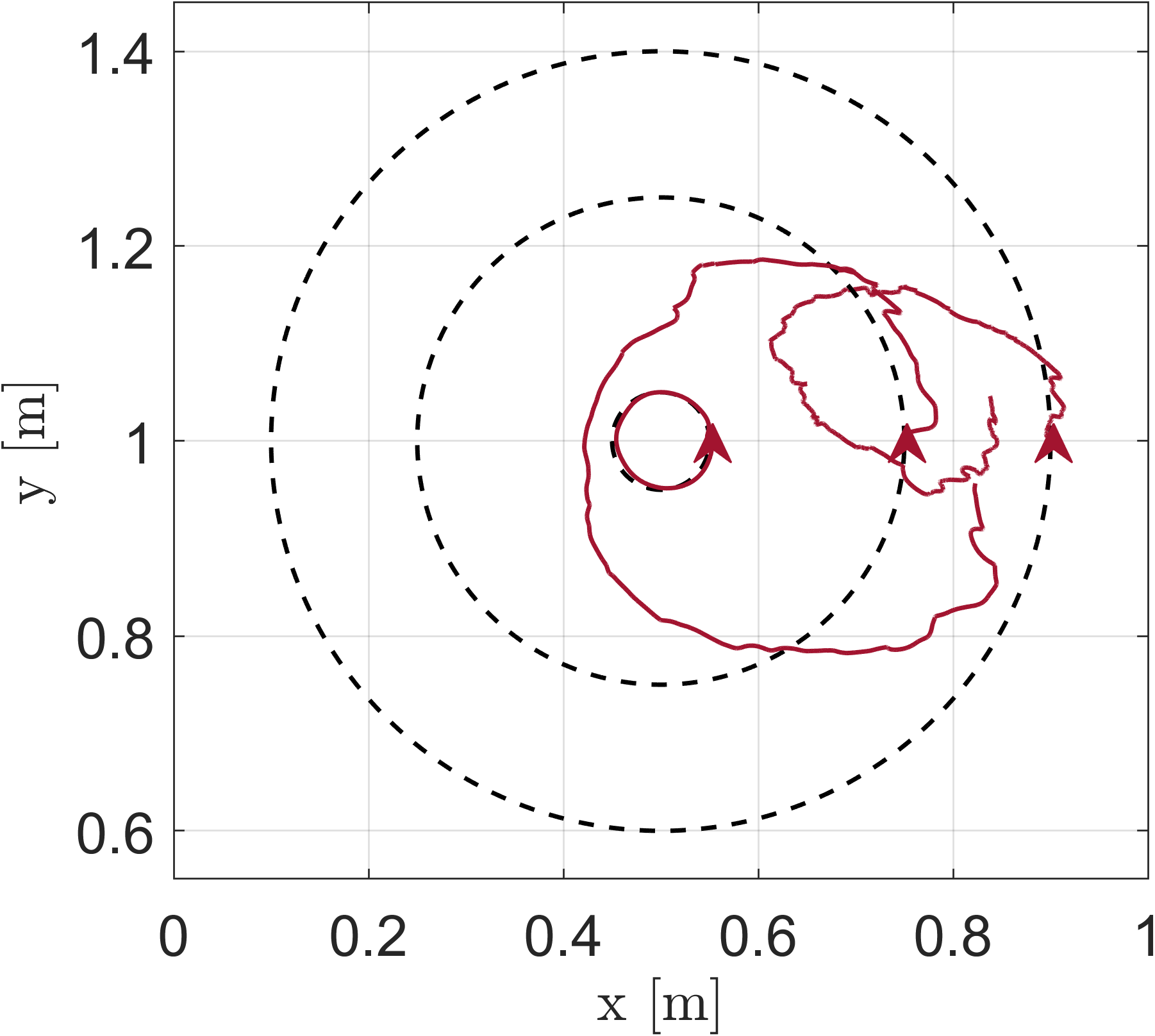}%
    \label{fig:DMDcTracking}}
    \hfil
    \subfloat[Bilinear]{\includegraphics[height=5.1cm]{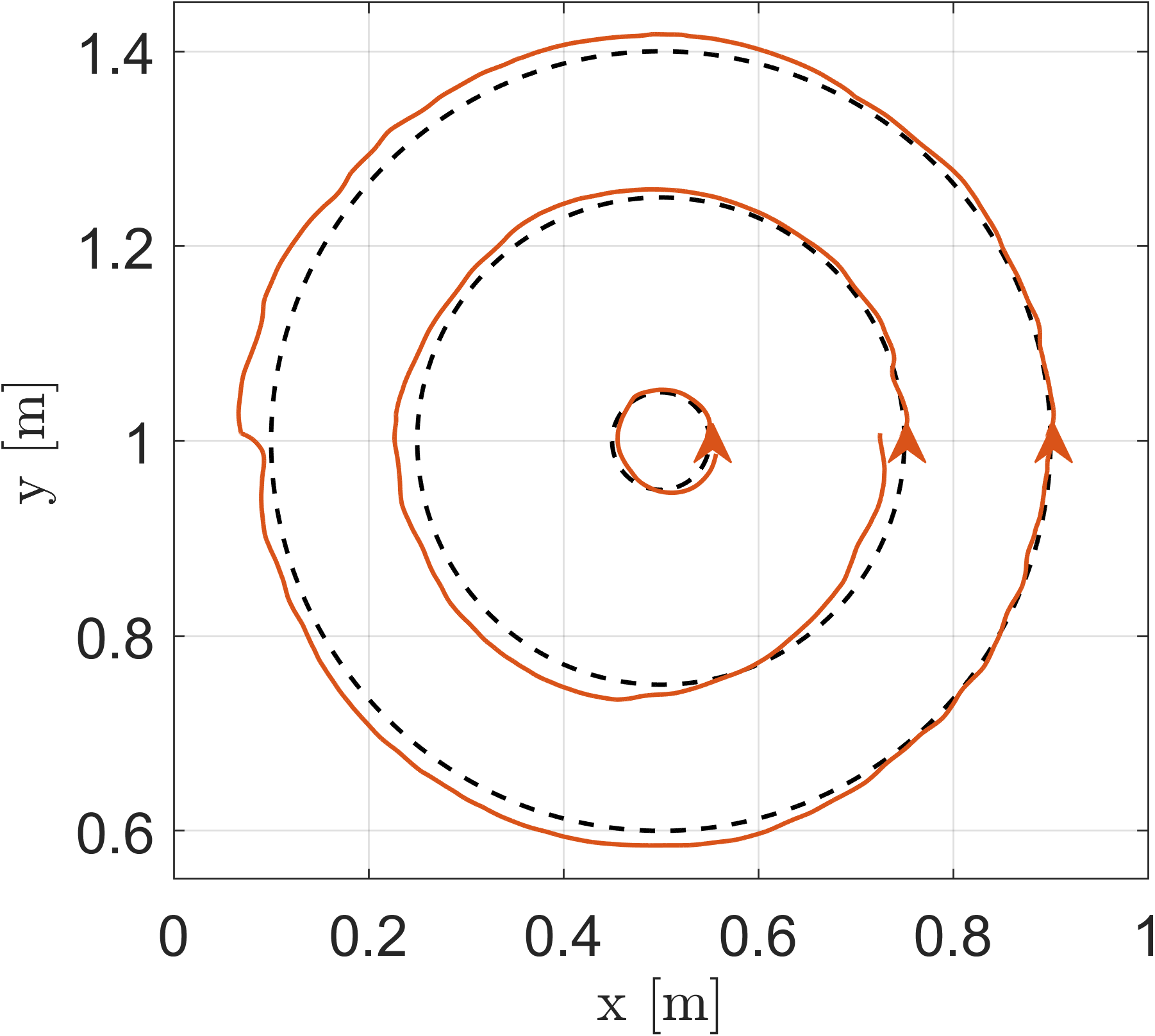}%
    \label{fig:BilinearTracking}}
    \caption{MPC control comparison of (a) Control-Coherent Koopman, (b) DMDc, and (c) Bilinear \cite{Bruder2021AdvantagesDynamicsb} model for circular trajectories with different radii. Tracking performance is summarized in Table \ref{tab:Comparison}.}%
    \label{fig:Comparison}%
\end{figure*}

Fig. \ref{fig:Configuration} shows the three trajectories used for evaluating the MPC control performance and their corresponding configuration rage. Fig.\ref{fig:CCKTracking} shows the MPC tracking accuracy of CCK. The tracking performance is satisfactory for all the three circular trajectories. In contrast, Fig.\ref{fig:DMDcTracking} shows \edits{the tracking performance of} DMDc which is only able to track the smallest circle. 

Following \cite{Bruder2021AdvantagesDynamicsb}, a bilinear model is constructed by using the same data and the same observables (with the same centers and dilation factors) as in the CCK and DMDc models. Namely, the $A$ matrix has the same dimensions as $A_{CCK}$ and $A_{DMDc}$. Fig.\ref{fig:BilinearTracking} shows the bilinear control performance. Although it tracks all the three trajectories, the tracking accuracy of the bilinear case is inferior compared to CCK, despite having a richer model. As the trajectory increases and the bilinear component of the model varies more significantly, the bilinear model yields significantly inferior results compared to CCK. See Table \ref{tab:Comparison}.

\begin{figure*}[!t]
    \centering
    \subfloat[]{\includegraphics[width=0.685\columnwidth]{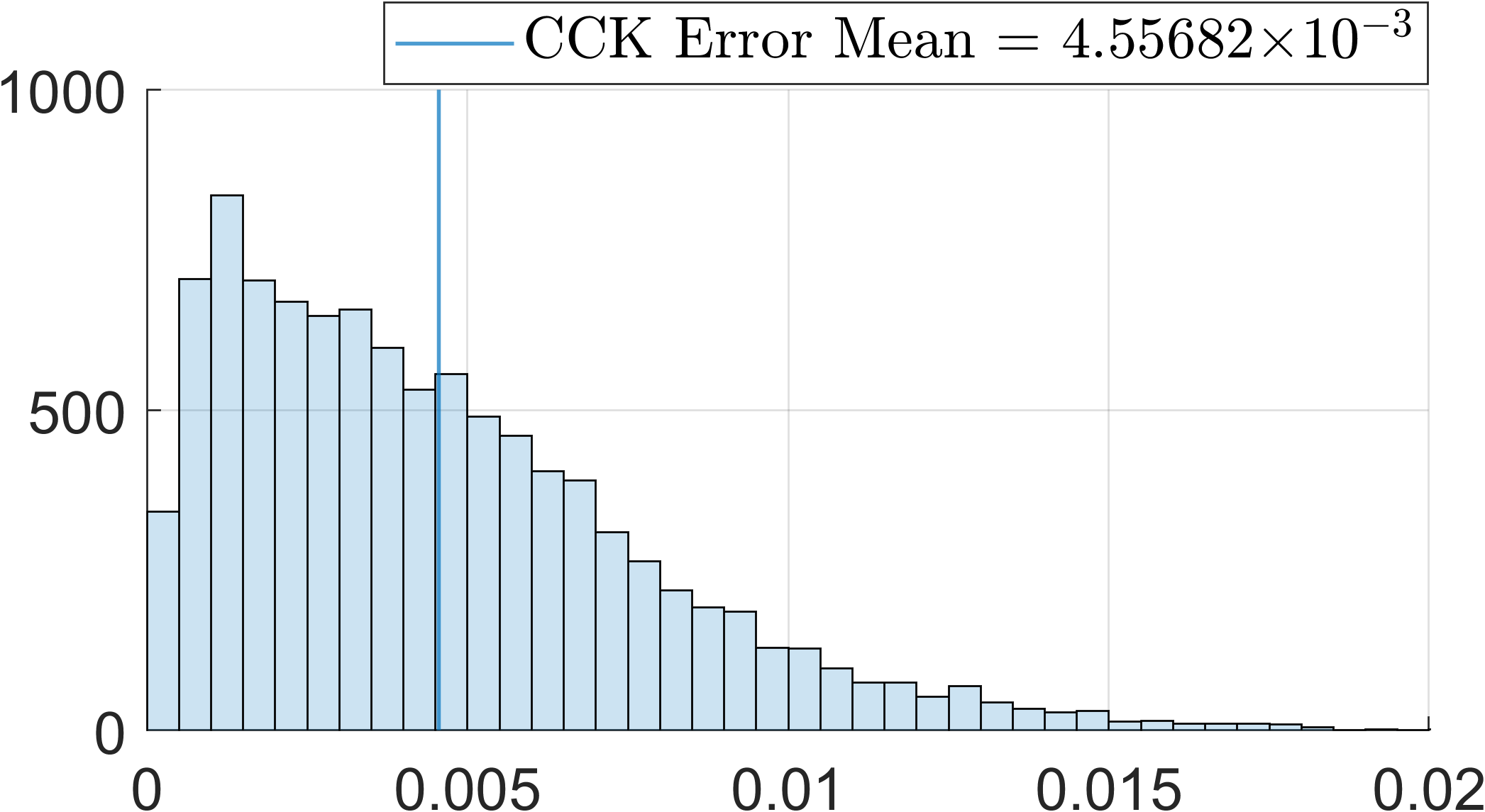}%
    \label{fig:ErrorDistCCK}}
    \hfill
    \subfloat[]{\includegraphics[width=0.685\columnwidth]{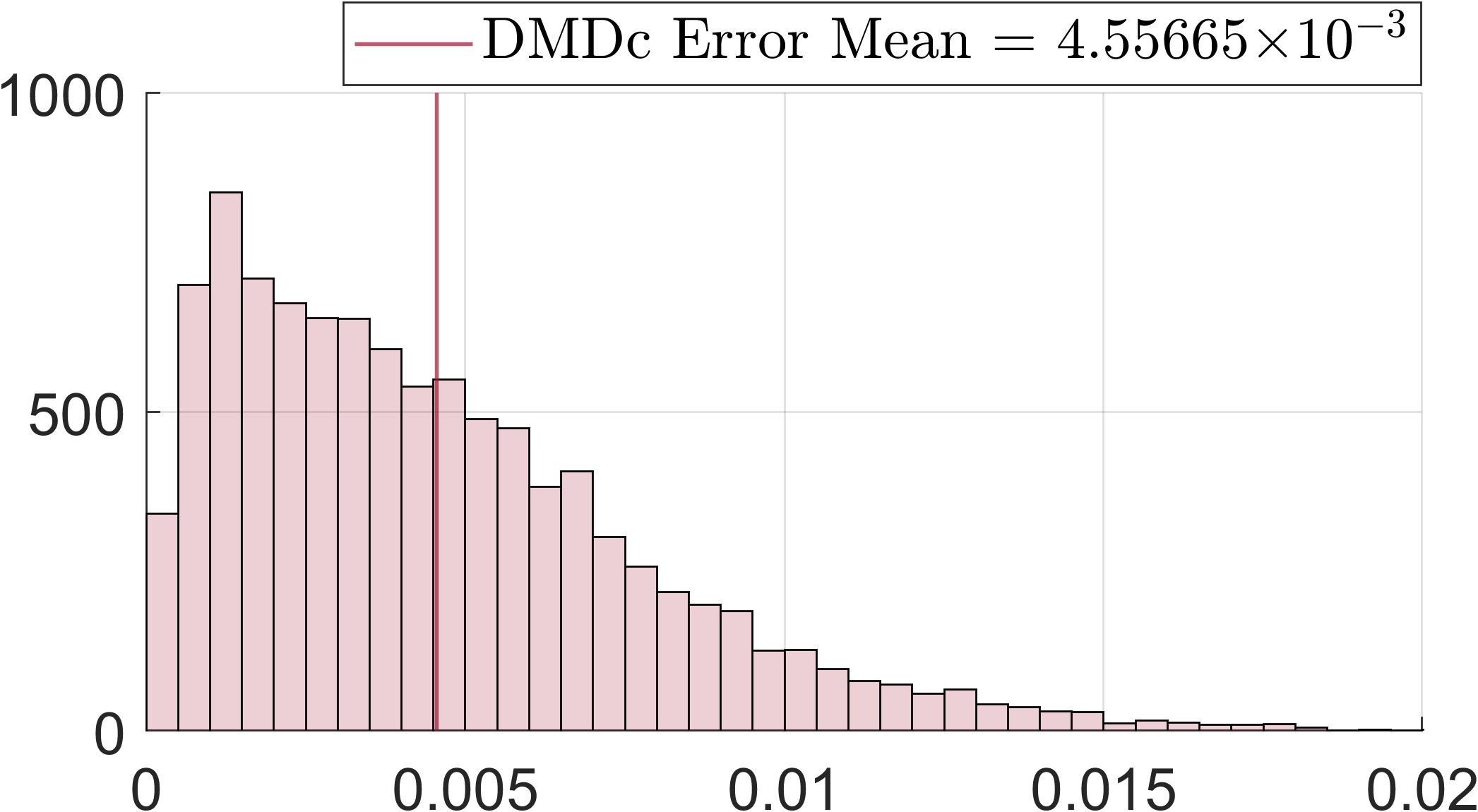}%
    \label{fig:ErrorDistDMD}}
    \hfill
    \subfloat[]{\includegraphics[width=0.685\columnwidth]{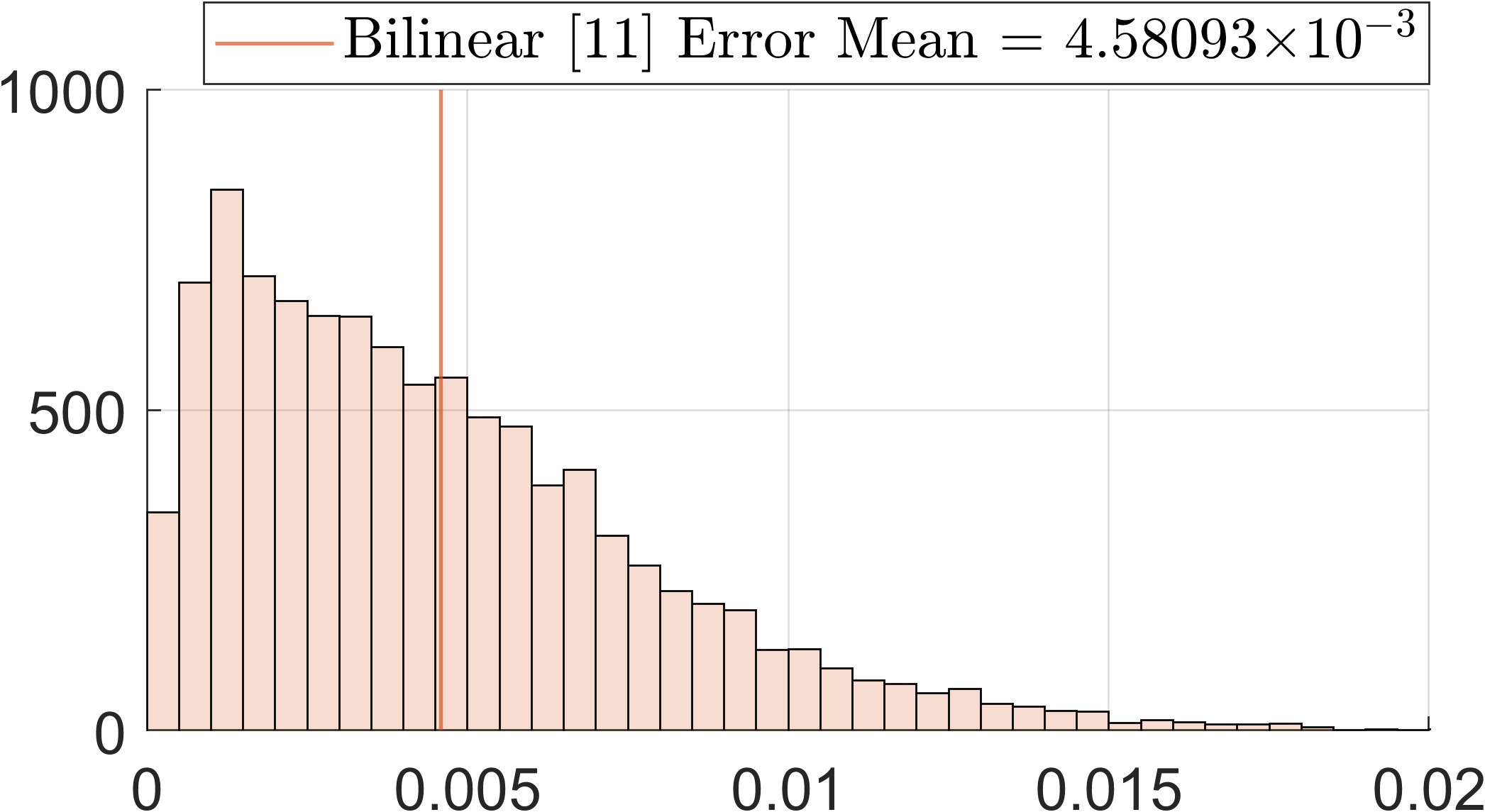}%
    \label{fig:ErrorDistBilinear}}
    \caption{Histogram comparison shows almost identical prediction accuracy between (a) Control-Coherent Koopman, (b) DMDc, and (c) Bilinear Koopman \cite{Bruder2021AdvantagesDynamicsb}.}
    \label{fig:Error}
\end{figure*}

The three plots in Fig. \ref{fig:Error} show the error histograms of one-step ahead state prediction of CCK, DMDc, and the bilinear model, respectively. Note that the prediction error histogram is essentially the same for the three methods. Nonetheless, the MPC \footnote{To prevent erratic behaviors in the DMDc simulation, a bound on the magnitude of the control effort, set at $20 \ N/m$, is incorporated into the MPC optimization.} tracking performance is strikingly different as summarized in Table \ref{tab:Comparison}.

\begin{table}[htbp]
\caption{Mean tracking error comparison between CCK, DMDc and Bilinear \cite{Bruder2021AdvantagesDynamicsb} model for trajectories with different radii}
\label{tab:Comparison}
\begin{center}
\begin{tabular}{|c|c|c|c|}
\hline
Radius [cm] & \begin{tabular}[c]{@{}c@{}} CCK Mean\\ Error [cm]\end{tabular} & \begin{tabular}[c]{@{}c@{}}DMDc Mean\\ Error [cm]\end{tabular} & \begin{tabular}[c]{@{}c@{}}Bilinear \cite{Bruder2021AdvantagesDynamicsb} \\ Mean Error [cm]\end{tabular} \\ \hline
5 & \textbf{0.76} & 1.06 & 0.95 \\ \hline
25 & \textbf{1.75} & 13.53 & 1.78 \\ \hline
40 & \textbf{1.70} & 37.74 & 2.33 \\ \hline
\end{tabular}
\end{center}
\end{table}
Looking into the poor performance of the standard DMDc as compared against CCK, the main difference between the two approaches comes from how the input matrix is constructed. While the input matrix of CCK has non-zero elements only in the block of actuator dynamics (\ref{eq:p_phi}), DMDc produces non-zero elements in both blocks. The $B_{DMDc}$ allows the control input, i.e. actuator torques, to change the joint angles directly and instantaneously, which is not coherent with the physical model. \edits{In the discretized model, t}he torques can only influence the transition of the velocities. When the MPC generates control signals based on the $B_{DMDc}$ and applies them to the plant, it performs poorly.
\begin{figure}[H]
    \centering  \includegraphics[width=0.69\columnwidth]{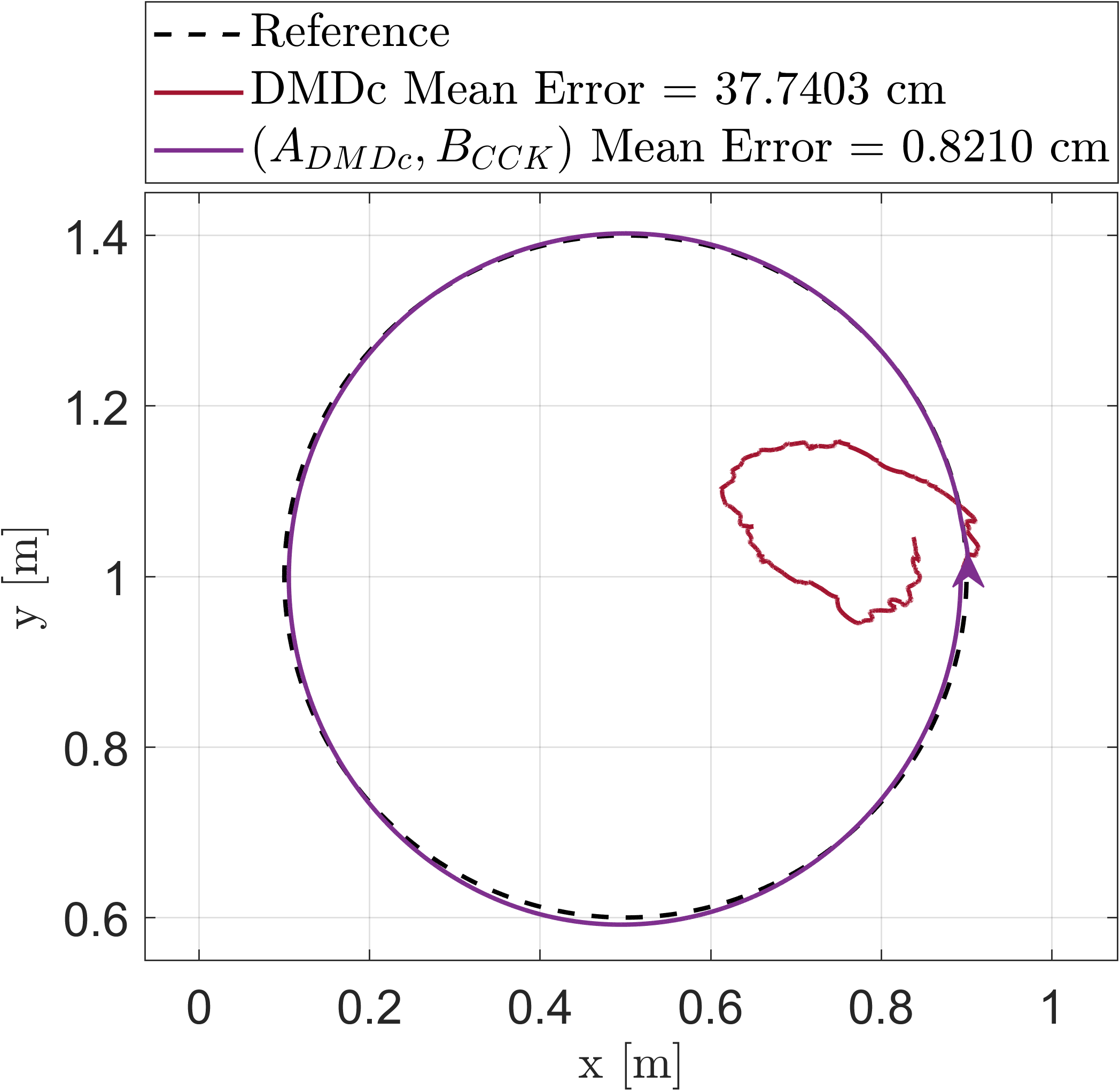}
    \caption{MPC control comparison of DMDc and a hybrid of $A_{DMDc}$ and $B_{CCK}$.}
    \label{fig:DMDc_modified}
\end{figure}

Although the discrepancy in prediction accuracy seems insignificant between the two, when applied to the MPC problem, it has a profound effect on the tracking performance. This is further confirmed with the additional numerical experiment in Fig. \ref{fig:DMDc_modified}. Here, MPC is run with the $A$ matrix from DMDc $(A_{DMDc})$ but with the input matrix from Control-Coherent
Koopman $(B_{CCK})$. The improvement on the tracking performance is significant as seen in the figure. This highlights the importance of using the coherent input matrix $(B_{CCK})$ rather than merely fitting training data. The MPC controller must not be misinformed with a questionable input matrix.

\bibliographystyle{ieeetr}

\end{document}